\documentclass[journal]{IEEEtran}

\usepackage[usenames]{color}
\usepackage{balance}
\usepackage{amsmath}
\usepackage{amssymb}
\usepackage{multirow}
\usepackage{cite}
\usepackage{ifpdf}
\usepackage{hyperref}

\ifCLASSINFOpdf
   \usepackage[pdftex]{graphicx}
   \graphicspath{{./pdf/}}
   \DeclareGraphicsExtensions{{.pdf}}
\else
   \usepackage[dvips]{graphicx}
   \graphicspath{{./eps/}}
   \DeclareGraphicsExtensions{{.eps}}
\fi
\begin{document}

\title{ {Exclusion and Guard Zones in DS-CDMA \\ Ad Hoc Networks}
\author{Don Torrieri, {\em Senior Member, IEEE}, and~Matthew C. Valenti, {\em Senior Member, IEEE}}
\thanks{Portions of this paper were published in the Proceedings of the IEEE Military Communications Conference (MILCOM), Orlando, FL, 2012.}
\thanks{M.C. Valenti's contribution was sponsored by the National Science Foundation under Award No. CNS-0750821 and by the United States Army Research Laboratory under Contract W911NF-10-0109.}
\thanks{D. ~Torrieri is with the US Army Research Laboratory, Adelphi, MD (email: don.j.torrieri.civ@mail.mil).}
\thanks{M.~C.~Valenti is with West Virginia University, Morgantown, WV, U.S.A. (email: valenti@ieee.org).} }
\maketitle

\begin{abstract}
The central issue in direct-sequence code-division multiple-access (DS-CDMA)
ad hoc networks is the prevention of a near-far problem. This paper
considers two types of guard zones that may be used to control the near-far
problem: a fundamental \emph{exclusion zone} and an additional \emph{CSMA
guard zone} that may be established by the carrier-sense multiple-access
(CSMA) protocol. In the exclusion zone, no mobiles are physically present,
modeling the minimum physical separation among mobiles that is always
present in actual networks. Potentially interfering mobiles beyond a
transmitting mobile's exclusion zone, but within its CSMA guard zone, are
deactivated by the protocol. This paper provides an analysis of DS-CSMA
networks with either or both types of guard zones. A network of finite
extent with a finite number of mobiles and uniform clustering as the spatial
distribution is modeled. The analysis applies a closed-form expression for the
outage probability in the presence of Nakagami fading, conditioned on the
network geometry. The
tradeoffs between exclusion zones and CSMA guard zones are explored for
DS-CDMA and unspread networks. The spreading factor and the guard-zone radius
provide design flexibility in achieving specified levels of average outage
probability and transmission capacity. The advantage of an exclusion zone
over a CSMA guard zone is that since the network is not thinned, the number
of active mobiles remains constant, and higher transmission capacities can
be achieved.
\end{abstract}

\begin{IEEEkeywords}
Spread-spectrum communication, Carrier-sense multiaccess, Mobile communications, Fading channels.
\end{IEEEkeywords}

\thispagestyle{empty}

\section{Introduction}

Direct-sequence code-division multiple-access (DS-CDMA) ad hoc networks are
realized by using direct-sequence spread-spectrum modulation while the
mobiles or nodes of multiple users simultaneously transmit signals in the
same frequency band. All signals use the entire allocated spectrum, but the
spreading sequences differ. DS-CDMA is advantageous for ad hoc networks
because it eliminates the need for any frequency or time-slot coordination,
imposes no sharp upper bound on the number of mobiles, directly benefits
from inactive terminals in the network, and is capable of efficiently
implementing bursty data traffic, intermittent voice signals, multibeamed
arrays, and reassignments to accommodate variable data rates. Furthermore,
DS-CDMA systems are inherently resistant to interference, interception, and
frequency-selective fading.

The central issue in DS-CDMA ad hoc networks is the prevention of a near-far
problem. The solution to the near-far problem in cellular networks is power
control \cite{torr}. However, the absence of a centralized control of an ad
hoc network renders any attempted power control local rather than pervasive.
As an alternative to power control, ad hoc networks typically use \emph{%
guard zones} surrounding mobiles to manage interference. This paper
considers two types of guard zones to control the near-far problem: a
fundamental \emph{exclusion zone} and an additional guard zone that may be
established by the carrier-sense multiple-access (CSMA) protocol; i.e., a 
\emph{CSMA guard zone}. In the exclusion zone surrounding a mobile, no other
mobiles are physically present.  The exclusion zone is used to model networks
for which there is a minimum physical separation among mobiles, for 
instance, when there is a one-to-one correspondence between mobiles and vehicles,
and a spacing is required to prevent collision. In the CSMA guard zone \cite{krunz}, 
\cite{alaw}, \cite{hasan}, which extends beyond the exclusion zone, other mobiles
may be present, but they are deactivated by the protocol. The CSMA guard
zone offers additional near-far protection beyond that offered by the
exclusion zone, but at the cost of reduced network transmission capacity, as
shown subsequently.  However, CSMA guard zones are useful for operating environments
that do not permit large exclusion zones, such as networks with low mobility or
a high density of mobile terminals.

In this paper, the network topology is modeled stochastically by specifying the 
spatial distribution of its mobiles.  A similar approach was taken by Hasan and 
Andrews \cite{hasan} to study guard zones, which like \cite{weber,win,gula,salb,haen,bacc2,ganti} 
leveraged tools from stochastic geometry \cite{stoy}, \cite{bacc1}.  However,
the existing literature makes the unrealistic assumption that the network is
infinite in extent with mobiles that are drawn from a Poisson point process.
A key weakness with the use of a Poisson point process is that it 
does not preclude an infinite number of mobiles from being located 
within a small region, which renders it inapplicable to the
study of guard zones.  This weakness limited \cite{hasan} to only place
a guard zone around the reference receiver.  In contrast, the network
model in the present paper allows a guard zone around \emph{every} mobile in the network,
which is possible by placing minimum-spacing constraints on the spatial model.  
Spatial constraints are studied in \cite{ganti} in the context of infinite networks, 
but the analysis is intractable except when operating at high signal-to-interference 
ratio (SIR).

The analysis in this paper builds upon a recent closed-form expression 
for the \emph{exact} outage probability \cite{torr3} at a reference receiver
conditioned on the network realization. The analysis allows for the arbitrary
placement of mobiles, arbitrary shadowing factors, fading with a Nakagami-m 
factor that can vary among the channels from the mobiles to the reference-receiver, 
thermal noise, and an arbitrary duty factor for each mobile.   Simulation
results in \cite{torr3} confirmed the exactness of the analysis. In the present paper,
we apply the analysis of \cite{torr3} to  obtain the \emph{spatially averaged} outage probability 
for networks that use guard zones.   The spatial averaging is obtained by taking the 
expected value of the outage probability with respect to the spatial distribution and
distribution of the shadowing.  While this expectation is not
generally analytically tractable when guard zones are present, a
simple and efficient Monte Carlo method can be used that involves the random
placement of mobiles and generation of shadowing factors.

The contributions of this paper are as follows.  A fundamental contribution of
this paper is that it presents an effective approach to analyzing finite networks 
with additional spatial constraints imposed. Such constraints could include 
a minimum physical separation between mobiles, and the deactivation of mobiles 
within a certain radius of an active mobile.  Using this analytical approach,
the impact of both types of guard zones on performance is studied.  The performance
is quantified by the spatially averaged outage probability, latency, and 
transmission capacity.  The numerical results in this paper show
the influence and relationship of essential parameters including the reference
receiver location, the spreading factor, the CSMA guard-zone radius, the exclusion-zone
radius, the transmission distance, and the number of potentially interfering mobiles.
The principal advantages of our analysis relative to ones based on stochastic geometry are
the accommodation of a network of arbitrary size and shape and an arbitrary
spatial distribution of mobiles, the ability to distinguish among the
performances of mobiles located anywhere in the interior of a network and
its outer edges, and the accommodation of arbitrary fading and shadowing
parameters for every path between two mobiles.

The remainder of the paper is organized as follows.  Section \ref{Section:GuardZones}
describes exclusion zones and CSMA guard zones, and Section \ref{Section:NetworkModel}
presents the network model. Section \ref{Section:Outage} reviews the conditional
outage probability, where the conditioning is on the network realization.
Examples of conditional outage probability both with and without CSMA guard
zones are given in Section \ref{Section:Examples}.  The conditioning is removed
in Section \ref{Section:SpatialAveraging} by averaging the outage
probability with respect to the shadowing and the spatial distribution 
of the mobiles.  Numerical results are given in Section \ref{Section:TC}. 
Finally, Section \ref{Section:Conclusions} concludes the paper.

\section{Guard Zones}

\label{Section:GuardZones}

The IEEE 802.11 standard, which is currently the predominant standard for ad
hoc networks, uses CSMA with collision avoidance 
in its medium-access control protocol \cite{krunz}, \cite{alaw}. The
implementation entails the exchange of request-to-send (RTS) and
clear-to-send (CTS) handshake packets between a transmitter and receiver
during their initial phase of communication that precedes the subsequent
data and acknowledgment packets. The receipt of the RTS/CTS packets with
sufficient power levels by nearby mobiles causes them to inhibit their own
transmissions, which would produce interference in the receiver of interest.
The transmission of separate CTS packets in addition to the RTS packets
decreases the possibility of subsequent signal collisions at the receiver
due to nearby hidden terminals that do not sense the RTS packets. Thus, the
RTS/CTS packets essentially establish guard zones surrounding a transmitter
and receiver, and hence prevent a near-far problem except during the initial
reception of an RTS packet. The interference at the receiver is restricted
to concurrent transmissions generated by mobiles outside the guard zones.
The fundamental disadvantage with this CSMA guard zone is that it inhibits
other concurrent transmissions within the zone, thereby reducing the network
transmission capacity.

A more fundamental guard zone, which is called the \textit{exclusion zone},
is based on the spacing that occurs in actual mobile networks. For instance,
when the radios are mounted on separate vehicles, there is a need for crash avoidance
by maintaining a minimum vehicle separation. A small exclusion zone
maintained by visual sightings exists in practical networks, but a more
reliable and extensive one can be established by equipping each mobile with
a global positioning system (GPS) and periodically broadcasting each
mobile's GPS coordinates. Mobiles that receive those messages could compare
their locations to those in the messages and alter their movements
accordingly. The major advantage of the exclusion zone compared with a CSMA
guard zone is that the exclusion zone prevents near-far problems at
receivers while not inhibiting any potential concurrent transmissions.
Another advantage of an exclusion zone is enhanced network connectivity
because of the inherent constraint on the clustering of mobiles.

When CSMA is used in a network, the CSMA guard zone will usually encompass
the exclusion zone. Although both zones may cover arbitrary regions, they
are modeled as circular regions in the subsequent examples for computational
convenience, and the region of the CSMA guard zone that lies outside the
exclusion zone is an annular ring. The existence of an annular ring enhances
the near-far protection at the cost of inhibiting potential concurrent
transmissions within the annular ring. The tradeoffs entailed in having the
CSMA guard zone are examined subsequently.

\section{Network Model}

\label{Section:NetworkModel} The network comprises a fixed number of mobiles
in a circular area with radius $r_\mathsf{net}$, although any arbitrary two-
or three-dimensional regions could be considered. A reference receiver is
located inside the circle, a reference transmitter $X_{0}$ is located at
distance $||X_{0}||$ from the reference receiver, and there are $M$
potentially interfering mobiles $X_{1},...,X_{M}.$ The variable $X_{i}$
represents both the $i^{th}$ mobile and its location, and $||X_{i}||$ is the
distance from the $i^{th}$ mobile to the reference receiver. Each mobile uses a single
omnidirectional antenna. The radii of the exclusion zone and the CSMA guard
zone are $r_\mathsf{ex}$ and $r_\mathsf{g},$ respectively.

The potentially interfering mobiles are uniformly distributed throughout the network
area outside the exclusion zones, according to a \textit{uniform clustering}
model. One by one, the location of each $X_i$ is drawn according to a
uniform distribution within the radius-$r_\mathsf{net}$ circle. However, if
an $X_i$ falls within the exclusion zone of a previously placed mobile, then
it has a new random location assigned to it as many times as necessary until
it falls outside all exclusion zones. Unlike Matern thinning \cite{card},
which silences mobiles without replacing them, uniform clustering maintains
a fixed number of potentially interfering mobiles. Setting the exclusion zone to $r_\mathsf{ex%
} = 0$ is equivalent to drawing the mobiles from a binomial point process.

Since there is no significant advantage to using short spreading sequences
in an asynchronous ad hoc network, long spreading sequences are assumed and
modeled as random binary sequences with chip duration $T_{c}$. The \emph{%
spreading factor} or \emph{processing gain} $G$ directly reduces the
interference power. The multiple-access interference is assumed to be
asynchronous, and the power from each interfering $X_{i}$ is further reduced
by the chip function $h(\tau _{i})$, which is a function of the chip waveform
and the timing offset $\tau _{i}$ of $X_{i}$'s spreading sequence relative
to that of the desired signal \cite{torr}. Since only timing offsets modulo-$%
T_{c}$ are relevant, $0\leq \tau _{i}<T_{c}$. In a network of quadriphase
direct-sequence systems, a multiple-access interference signal with power $%
\mathcal{I}_{i}$ before despreading is reduced after despreading to the
power level $\mathcal{I}_{i}h(\tau _{i})/G,$ where \cite{torr}, \cite{torr2} 
\begin{equation}
h(\tau _{i})=\frac{1}{T_{c}^{2}}\left[ R_{\psi }^{2}(\tau _{i})+R_{\psi
}^{2}(T_{c}-\tau _{i})\right] 
\end{equation}%
and $R_{\psi }(\tau _{i})$ is the partial autocorrelation for the normalized
chip waveform. Thus, the interference power is effectively reduced by the
factor $G_{i}=G/h(\tau _{i}).$ Assuming a rectangular chip waveform and that 
$\tau _{i}$ has a uniform distribution over [0, $T_{c}],$ 
the expected value of $h(\tau _{i})$ is 2/3.

After the despreading, the power of $X_{i}$'s signal at the reference receiver
is 
\begin{equation}
\rho_{i}= \tilde{P}_{i}g_{i}10^{\xi_{i}/10}f\left( ||X_{i}|| \right)
\label{power}
\end{equation}
where $\tilde{P}_{i}$ is the received power at a reference distance $d_0$
(assumed to be sufficiently far that the signals are in the far field) after
despreading when fading and shadowing are absent, $g_{i}$ is the power gain
due to fading, $\xi_{i}$ is a shadowing factor, and $f(\cdot)$ is a
path-loss function. The path-loss function is expressed as the power law 
\begin{equation}
f\left( d\right) =\left( \frac{d}{d_0}\right) ^{-\alpha}\hspace{-0.45cm}, \,
\, \text{ \ }d\geq d_0  \label{pathloss}
\end{equation}
where $\alpha\geq2$ is the path-loss exponent. It is assumed that $r_\mathsf{%
ex} \geq d_0.$ The \{$g_{i}\}$ are independent with unit-mean, but are not
necessarily identically distributed; i.e., the channels from the different $%
\{X_{i}\}$ to the reference receiver may undergo fading with different
distributions. For analytical tractability and close agreement with measured
fading statistics, Nakagami fading is assumed, and $g_{i}=a_{i}^{2}$, where $%
a_{i}$ is Nakagami with parameter $m_{i}$. When the channel between $X_{i}$
and the reference receiver undergoes Rayleigh fading, $m_{i}=1$ and the
corresponding $g_{i}$ is exponentially distributed. In the presence of
log-normal shadowing, the $\{\xi_{i}\}$ are independent zero-mean Gaussian
with variance $\sigma_{s}^{2}$. For ease of exposition, it is assumed that
the shadowing variance is the same for the entire network, but the results
may be easily generalized to allow for different shadowing variances over
parts of the network. In the absence of shadowing, $\xi_{i}=0$.

It is assumed that the \{$g_{i}\}$ remain fixed for the duration of a time
interval but vary independently from interval to interval (block fading).
With probability $p_{i}$, the $i^{th}$ mobile transmits in the same time
interval as the desired signal. The $\{p_{i}\}$ can be used to model
voice-activity factors, controlled silence, or failed link transmissions and
the resulting retransmission attempts. The $\{p_{i}\}$ need not be the same;
for instance, CSMA protocols can be modeled by setting $p_{i}=0$ when a
mobile lies within the CSMA guard zone of another active mobile, which is
equivalent to implementing Matern thinning within the annular ring
corresponding to that guard zone.

The instantaneous signal-to-interference-and-noise ratio (SINR) at the
reference receiver is given by: 
\begin{equation}
\gamma=\frac{\rho_{0}}{\displaystyle{\mathcal{N}}+\sum_{i=1}^{M}I_{i}\rho_{i}%
}  \label{SINR1}
\end{equation}
where $\rho_{0}$ is the received power of the desired signal, $\mathcal{N}$
is the noise power, and the indicator $I_{i}$ is a Bernoulli random variable
with probability $P[I_{i}=1]=p_{i}$ and $P[I_{i}=0]=1-p_{i}$.

Since the despreading does not significantly affect the desired-signal power, the
substitution of (\ref{power}) and (\ref{pathloss}) into (\ref{SINR1}) yields 
\begin{equation}
\gamma=\frac{g_{0}\Omega_{0}}{\displaystyle\Gamma^{-1}+%
\sum_{i=1}^{M}I_{i}g_{i}\Omega_{i}}  \label{SINR2}
\end{equation}
where 
\begin{equation}
\Omega_{i}=%
\begin{cases}
10^{\xi_{0}/10}||X_{0}||^{-\alpha} & i=0 \\ 
\displaystyle\frac{{P}_{i}}{G_i P_{0}}10^{\xi_{i}/10}||X_{i}||^{-\alpha} & 
i>0%
\end{cases}
\label{eqn:omega}
\end{equation}
is the normalized power of $X_{i}$, ${P}_{i}$ is the received power at the
reference distance $d_0$ before despreading when fading and shadowing are
absent, and $\Gamma=d_0^{\alpha}P_{0}/\mathcal{N}$ is the SNR when the
reference transmitter is at unit distance from the reference receiver and
fading and shadowing are absent.

\section{Outage Probability}
\label{Section:Outage} 
The {\em outage probability} quantifies the likelihood that the noise and interference will be too severe for useful communications.  Outage probability is defined with respect to an SINR threshold $\beta$, which represents  the minimum SINR required for reliable reception.  In general, the value of $\beta$ depends on the choice of coding and modulation.   An \emph{outage} occurs when the
SINR falls below $\beta $.

In \cite{torr3}, closed-form expressions were found for the outage probability conditioned on the particular network geometry and shadowing factors.  Let $\boldsymbol{\Omega }=\{\Omega _{0},...,\Omega _{M}\}$
represent the set of normalized powers.  Conditioning on $\boldsymbol{\Omega }$, the
\emph{outage probability} is 
\begin{equation}
\epsilon =P\left[ \gamma \leq \beta \big|\boldsymbol{\Omega }\right] .
\label{Equation:Outage1}
\end{equation}%


Restricting the Nakagami parameter $m_{0}$ of
the channel between the reference transmitter and receiver to be
integer-valued, the outage probability conditioned on $\boldsymbol{\Omega}$
is found in \cite{torr3} to be 
\begin{eqnarray}
\epsilon
& = &
1 - e^{-\beta_{0}z}\sum_{s=0}^{m_{0}-1}{\left( \beta_{0}z\right) }%
^{s}\sum_{t=0}^{s}\frac{z^{-t}H_{t}(\boldsymbol{\Psi})}{(s-t)!}
\label{NakagamiCond}
\end{eqnarray}
where $\beta_{0}=\beta m_{0}/\Omega_{0}$, 
\begin{align}
\Psi_{i} & =\left( \beta_{0}\frac{\Omega_{i}}{m_{i}}+1\right) ^{-1}\hspace{%
-0.5cm},\hspace{1cm}\mbox{for $i=\{1,...,M\}$, }  \label{Psi} \\
H_{t}(\boldsymbol{\Psi}) & =\mathop{ \sum_{\ell_i \geq 0}}%
_{\sum_{i=0}^{M}\ell_{i}=t}\prod_{i=1}^{M}{\mathsf{G}}_{\ell_{i}}(\Psi_{i}),
\label{Hfunc}
\end{align}
the summation in (\ref{Hfunc}) is over all sets of indices that sum to $t$,
and \vspace{-0.35cm} 
\begin{equation}
\mathsf{G}_{\ell}(\Psi_{i})=%
\begin{cases}
1-p_{i}(1-\Psi_{i}^{m_{i}}) & \mbox{for $\ell=0$} \\ 
\frac{p_{i}\Gamma(\ell+m_{i})}{\ell!\Gamma(m_{i})}\left( \frac{\Omega_{i}}{%
m_{i}}\right) ^{\ell}\Psi_{i}^{m_{i}+\ell} & \mbox{for $\ell>0$.}%
\end{cases}
\label{Gfunc}
\end{equation}
The proof is found in Section III of \cite{torr3}.

\section{Examples}

\begin{figure}[t]
\centering
\vspace{-0.1cm} \includegraphics[width=7.5cm]{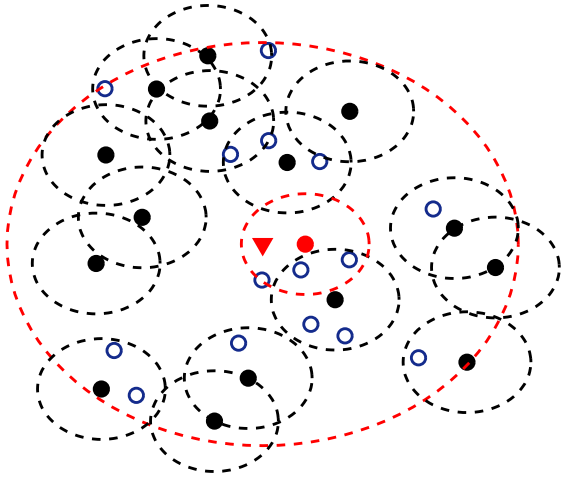} \vspace{-0.25cm%
}
\caption{ Example network realization. The reference receiver (indicated by
the {\color{red} $\blacktriangledown$}) is placed at the origin, and the
reference transmitter is at $X_{0}=1/6$ (indicated by the {\color{red} $%
\bullet$} to its right). $M=30$ mobiles are placed according to the uniform
clustering model, each with an exclusion zone (not shown) of radius $r_%
\mathsf{ex}=1/12$. Active mobiles are indicated by filled circles, and
deactivated mobiles are indicated by unfilled circles. A guard zone of
radius $r_\mathsf{g}=1/4$ surrounds each active mobile, as depicted by
dashed circles. When CSMA guard zones are used, the other mobiles within the
guard zone of an active mobile are deactivated. \protect\vspace{-0.65cm} }
\label{Figure:ExampleNetwork}
\end{figure}

\label{Section:Examples} In the examples, all distances are normalized to
the network radius so that $r_{\mathsf{net}}=1$, and the reference receiver
is located at the center of the network unless otherwise stated. As shown in
Fig. \ref{Figure:ExampleNetwork}, the reference transmitter is at coordinate 
$X_{0}=1/6$ and $M=30$ potentially interfering mobiles are placed according
to a uniform clustering model with a radius $r_{\mathsf{ex}}=1/12$ exclusion
zone surrounding each mobile. This radius is small enough that the exclusion
zones do not significantly impede the movements of mobiles. Once the mobile
locations $\{X_{i}\}$ are realized, the $\{\Omega _{i}\}$ are determined by
assuming a path-loss exponent $\alpha =3.5$ and a common transmit power ($%
{P}_{i}/P_{0}=1$ for all $i$). Both unshadowed and shadowed environments are
considered. 
Although the model permits nonidentical $G_{i}$, it is assumed that $G_{i}$
is a constant equal to $G_{\mathsf{e}}$ (the \emph{effective} processing
gain) for all interference signals. Both spread and unspread systems are
considered, with $G_{\mathsf{e}}=1$ for the unspread system and $G_{\mathsf{e%
}}=48$ for the spread system, corresponding to a typical direct-sequence
waveform with $G=32$ and $h(\tau _{i})=2/3$. Although the model permits
nonidentical $p_{i}$ in the range $\left[ 0,1\right] ,$ the value of $%
p_{i}=0.5$ for all active $X_{i}$ is chosen as an example, corresponding to 
a half-duplex mobile terminal with a full input buffer and a symmetric data transmission
rate to its peer terminal.
The Nakagami parameters are $m_{0}=3$ for the
desired signal and $m_{i}=1$ for the interfering mobiles (i.e., Rayleigh
fading). Using two different Nakagami factors (i.e., \emph{mixed} fading) is
justified by the fact that the reference transmitter is usually within the
line-of-sight of the reference receiver while the interfering mobiles are not
in the line-of-sight.
Therefore, the interference signals are subject to more severe fading. The
SINR threshold is $\beta =0$ dB, which corresponds to the unconstrained AWGN
capacity limit for a rate-$1$ channel code, assuming complex-valued inputs.

\begin{figure}[t]
\centering
\vspace{-0.1cm} \includegraphics[width=8.75cm]{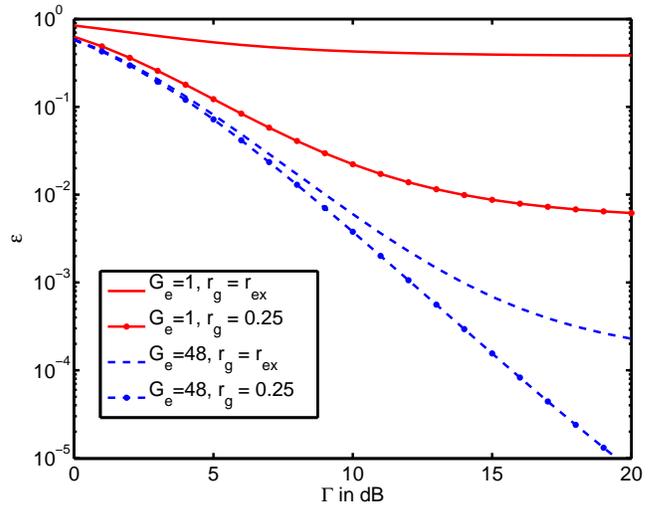} \vspace{%
-0.8cm}
\caption{ Outage probability as a function of SNR, conditioned on the
network shown in Fig. 1 with mixed fading, no shadowing, and $r_{ex}=1/12$.
Dashed lines are for spreading ($G_{\mathsf{e}}=48$), and solid lines are
for no spreading. Curves marked with $\bullet $ indicate performance with a CSMA
guard zone of radius $r_{\mathsf{g}}=1/4$, and curves without markers
indicate performance without CSMA guard zones. \protect\vspace{-0.65cm} }
\label{Figure:ConditionalOutage}
\end{figure}

\textbf{Example \#1.} Suppose that CSMA is not used, and therefore there is
no guard zone beyond the fundamental exclusion zone; i.e., $r_\mathsf{g}=r_%
\mathsf{ex}$. All 30 potentially interfering mobiles in Fig. \ref%
{Figure:ExampleNetwork} remain active and contribute to the overall
interference at the reference receiver. The outage probability for this
network is shown in Fig. \ref{Figure:ConditionalOutage} by the two curves
without markers, corresponding to the unspread ($G_\mathsf{e}=1$) and
direct-sequence spread ($G_\mathsf{e}=48$) networks. Without direct-sequence
spreading, the outage probability is quite high, for instance $\epsilon=0.4$
at $\Gamma=20$ dB. Spreading reduces the outage probability by about three
orders of magnitude at high SNR, although this comes at the cost of
increased required bandwidth.

\textbf{Example \#2.} The outage probability can be reduced by using CSMA to
impose guard zones beyond the fundamental exclusion zone. Suppose that a
guard zone of radius $r_\mathsf{g}=1/4$ is used, as shown in Fig. \ref%
{Figure:ExampleNetwork}. Potentially interfering mobiles are deactivated
according the following procedure, which is equivalent to Matern thinning.
First, the reference transmitter $X_{0}$ is activated. Next, each
potentially interfering mobile is considered in the order it was placed. For
each mobile, a check is made to see if it is in the guard zone of a
previously placed active mobile. Since mobiles are indexed according to the
order of placement, $X_{1}$ is first considered for possible deactivation;
if it lies in the guard zone of $X_{0}$, it is deactivated, and otherwise it
is activated. The process repeats for each subsequent $X_{i}$, deactivating
it if it falls within the guard zone of any active $X_{j}$, $j<i$, or
otherwise activating it.


In Fig. \ref{Figure:ExampleNetwork}, active mobiles are indicated by filled
circles and deactivated mobiles are indicated by unfilled circles. The guard
zone around each active mobile is indicated by a dashed circle of radius $r_%
\mathsf{g}$. The reference receiver has not been assigned a CSMA guard zone,
which reflects the fact that it has none while it is receiving the initial
RTS. In the given example, 15 mobiles have been deactivated, and the
remaining 15 mobiles remain active. The outage probability of the network
with deactivated mobiles is shown in Fig. \ref{Figure:ConditionalOutage} for 
$G_\mathsf{e}=\{1,48\}$ by the two curves with markers. The performance of
the unspread network improves dramatically with a guard zone, being reduced
by two orders of magnitude at high SNR. The DS-CDMA network, which already
had superior performance without CSMA, has improved performance, but the
improvement only becomes significant at a high SNR. 

\section{Spatial Averaging}

\label{Section:SpatialAveraging}

Because it is conditioned on ${\boldsymbol{\Omega }}$, the outage
probability often varies significantly from one network realization to the next. The
conditioning on ${\boldsymbol{\Omega }}$ can be removed by averaging the
conditional outage probability $\epsilon$ with respect to the network geometry. This averaging can
be done analytically only under certain limitations \cite{torr3}. For more
general cases of interest, the outage probability can be estimated through
Monte Carlo simulation by generating many different ${\boldsymbol{\Omega }}$%
, computing the outage probability of each, and taking the numerical
average. 

Suppose that $N$ networks are generated, and let $\epsilon_i$
be the outage probability of the $i^{th}$ network, whose vector of 
normalized powers is expressed as $\boldsymbol{\Omega }_{i}$.
The {\em average} outage probability may be found by taking the average:
\begin{eqnarray}
\bar{\epsilon} & = & 
{\mathbb E} \left( \epsilon \right)  
 = 
\frac{1}{N}\sum_{i=1}^{N} \epsilon_i.  
\label{Equation:MC}
\end{eqnarray}
Generating each ${\boldsymbol{\Omega }_i}$ involves not only placing
the mobiles according to the uniform clustering model, but also realizing
the shadowing and deactivating mobiles that lie within the CSMA guard zones
of the active mobiles.

\begin{figure}[t]
\centering
\vspace{-0.1cm} \includegraphics[width=8.75cm]{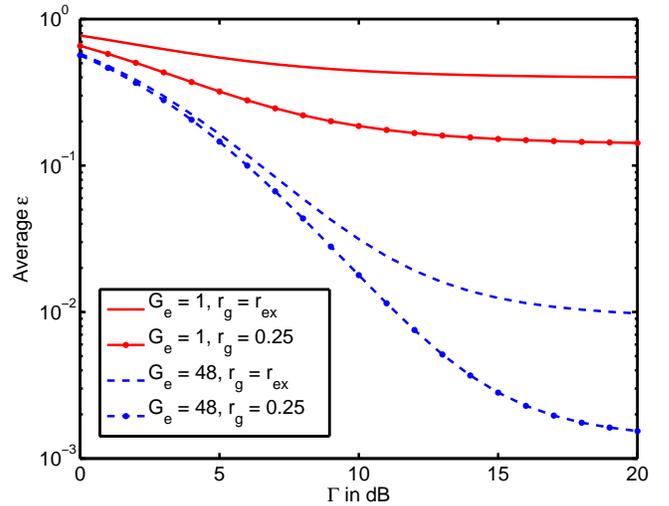} \vspace{-0.6cm}
\caption{ Outage probability in mixed fading as a function of SNR averaged
over $10,000$ networks drawn under the same conditions as the one shown in
Fig. 1. For each network, shadowing with $\protect\sigma _{s}=8$ dB is
applied and $r_{\mathsf{ex}}=1/12$. \ \protect\vspace{-0.4cm}}
\label{Figure:AverageOutage}
\end{figure}

As an example, Fig. \ref{Figure:AverageOutage} shows the average outage
probability computed over a set of $N=10,000$ networks, each generated the
same way as the Examples given in Section \ref{Section:Examples}. For $r_{%
\mathsf{g}}=r_{\mathsf{ex}}$, each network was generated by placing $M=30$
mobiles with exclusion zone $r_{\mathsf{ex}}=1/12$ according to a uniform
clustering model. For $r_{\mathsf{g}}=1/4$, the same set of networks was
used that was already generated with $r_{\mathsf{ex}}=1/12$, but in each
network, potentially interfering mobiles were deactivated if they were in
the guard zone of an active mobile. Log-normal shadowing with $\sigma _{s}=8$
dB was applied, and all other system parameters have the same values as in
Section \ref{Section:Examples}.

\section{Performance Analysis}
\label{Section:TC} 

This section uses the analytical tools developed in the previous sections
to investigate the relationship among essential parameters, including 
the CSMA guard-zone radius, the exclusion-zone radius, the spreading factor, 
the number of potentially interfering mobiles, the reference-receiver location, 
and the transmission distance.  Results are quantified by the spatially averaged
outage probability, transmission capacity, and average latency.  

Throughout this section, it is assumed that the SNR is $\Gamma =10$ dB, 
all channels undergo mixed fading ($m_{0}=3 $ and $m_{i}=1,i\geq 1$) 
with log-normal shadowing ($\sigma _{s}=8$ dB), the SINR threshold is $\beta =0$ dB,
and $p_{i}=0.5$ for the active mobiles.  The spatially averaged outage
probability is computed by averaging over $N=10,000$ network realizations.
Unless otherwise specified, the path-loss exponent is set to $\alpha =3.5$,
and the reference receiver is at the center of the network.

\subsection{Transmission Capacity}
While the outage probability is improved with CSMA due to
the deactivation of potentially interfering mobiles, the overall network
becomes less efficient due to the suppression of transmissions. The network
efficiency can be quantified by the \emph{transmission capacity} (TC) \cite%
{weber}: 
\begin{equation}
\tau=(1-\bar{\epsilon})\lambda b
\end{equation}
where $\lambda$ is the network density, which is the number of active
mobiles per unit area, and $b$ is the link throughput in the absence of an
outage, in units of bps. The TC represents the network throughput per unit
area. Increasing the size of the guard zone generally reduces TC due to
fewer simultaneous transmissions.

For a given value of $M$, the density without a CSMA guard zone remains
fixed since all mobiles remain active. However, with a CSMA guard zone, the
number of potentially interfering mobiles is random with a value that
depends on the value of $r_\mathsf{g}$, the locations of the mobiles, and
their order of placement, which affects how they are deactivated. As with
outage probability, Monte Carlo simulation can be used to estimate TC.

\subsection{Latency}

Assume that a selective-repeat hybrid ARQ protocol is used.  For such a protocol, the receiver attempts to decode the received packet, and signals back a positive acknowledgement (ACK) if it is correctly decoded or a negative acknowledgement (NACK) if it is not.  By setting $\beta$ to the minimum SINR required to correctly decode the packet, a retransmission will be required with probability $\epsilon$.  Assuming that type-I hybrid ARQ is used, the receiver flushes its memory and will attempt to independently decode each retransmitted packet.

Let $T_s$ be the ARQ time-slot duration, and define {\em latency} to be the average time from when a packet is sent to when it is first successfully decoded.  Let $N_\mathsf{ARQ}$ be the number of slots between the start of the initial transmission and the start of the retransmission, so that if a packet sent in time  slot $n$ fails, then it is retransmitted in slot $n+N_\mathsf{ARQ}$.   For such a protocol, the average probability of success in the first slot is $(1-\bar{\epsilon})$, and the average probability of 
success in slot $1+kN_\mathsf{ARQ}$ is $\bar{\epsilon}^{k}(1-\bar{\epsilon})$.  Taking the expected value of the number of 
slots before success and multiplying by the slot duration, the average latency $D$ is found to be
\begin{eqnarray}
D
& = &
T_s \left[ 
\frac{N_\mathsf{ARQ}}{1-\bar{\epsilon}}
- (1-\bar{\epsilon})(N_\mathsf{ARQ}-1)
\right].
\label{eqn:delay}
\end{eqnarray}


\subsection{Effect of $r_\mathsf{ex}$ and $r_\mathsf{g}$}

To investigate the influence of the exclusion-zone radius $r_{\mathsf{ex}}$ 
and guard-zone radius $r_{\mathsf{g}}$ on the network performance,  
the outage probability, transmission capacity, and latency were determined 
over a range of $r_{\mathsf{ex}}$ 
and $r_{\mathsf{g}}$.  To remove the dependence on the transmitter-receiver separation,
the guard and exclusion zones are normalized with respect to $||X_0||$.   The normalized
guard-zone radius was varied over $1/2 \leq r_{\mathsf{g}}/||X_0|| \leq 3$,
and three representative values of the normalized exclusion-zone radius were selected:  
$r_{\mathsf{ex}}/||X_0||=\{1/4,1/2,3/4\}$.  Both unspread and spread ($G_e = 48$)
networks were considered.  The number of potentially interfering mobiles
was set to $M=30$ and $r_\mathsf{net}/||X_0|| = 6$. 

\begin{figure}[t]
\centering
\vspace{-0.1cm} \includegraphics[width=8.75cm]{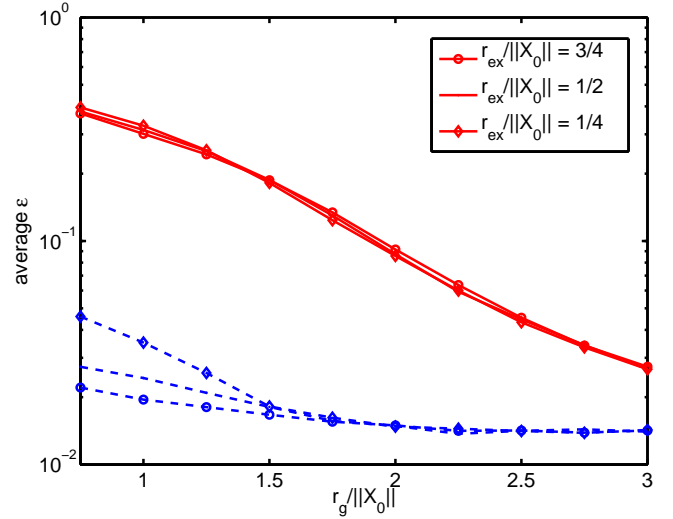} \vspace{-0.65cm}%
\vspace{-0.1cm}
\caption{Outage probability as a function of $r_{\mathsf{g}}/||X_0||$ for several
values of $r_{\mathsf{ex}}/||X_0||$ with $M=30$ and $r_\mathsf{net}/||X_0|| = 6$. 
Dashed lines are for spreading ($G_{\mathsf{e}}=48$), and solid lines are 
without ($G_{\mathsf{e}}=1$). }
\vspace{-0.2cm}
\label{Figure:OutageRg}
\end{figure}

Fig. \ref{Figure:OutageRg} shows the spatially averaged outage probability.
For the unspread system, the outage probability
is insensitive to the exclusion-zone radius but very sensitive to the guard-zone
radius.  This underscores the importance of a guard zone for an unspread network.
Except when the guard-zone radius is relatively small, the outage probability of the spread
system is insensitive to both the exclusion-zone and guard-zone radii.

\begin{figure}[t]
\centering
\vspace{-0.1cm} \includegraphics[width=8.75cm]{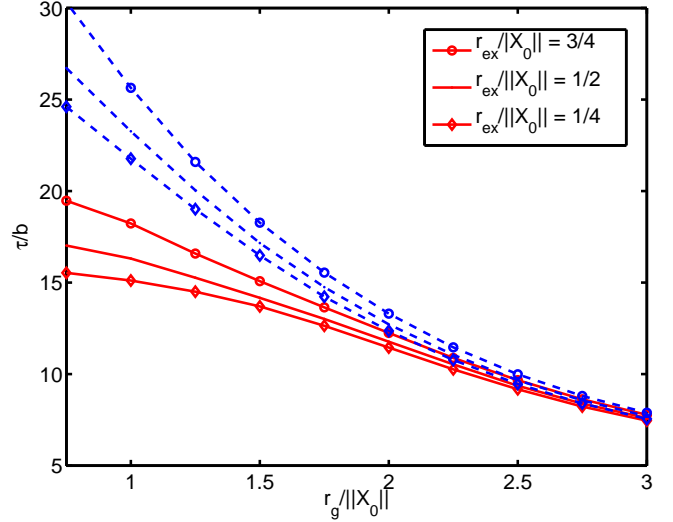} \vspace{-0.65cm}%
\vspace{-0.1cm}
\caption{Transmission capacity as a function of $r_{\mathsf{g}}/||X_0||$ for several
values of $r_{\mathsf{ex}}/||X_0||$ with $M=30$ and $r_\mathsf{net}/||X_0|| = 6$.
Dashed lines are for spreading ($G_{\mathsf{e}%
}=48$), and solid lines are for no spreading ($G_{\mathsf{e}}=1$).   }
\vspace{-0.1cm}
\label{Figure:TCvRg}
\end{figure}

Fig. \ref{Figure:TCvRg} shows the transmission capacity as a function of the guard-zone radius. 
Although the outage probability at the reference receiver in an unspread network is insensitive 
to $r_{\mathsf{ex}}$, Fig. \ref{Figure:TCvRg} shows that the TC is sensitive to $r_{\mathsf{ex}}$, especially
at low $r_{\mathsf{g}}$.  As $r_{\mathsf{ex}}$ increases, there are fewer nearby interfering
mobiles that get deactivated by the guard zone.  Thus, more mobiles remain active, and
the transmission capacity increases even though the outage probability remains fixed.
A similar behavior is seen for the spread network, which for small $r_{\mathsf{g}}$ has a significantly 
higher TC than the unspread network due to the lower outage probability. The main limitation to 
increasing $r_{\mathsf{ex}}$ is that it must  be small enough that there is no significant 
impediment to the movements of mobiles.  For both the spread and unspread
network, the TC diminishes quickly with increasing $r_{\mathsf{g}}$, and at high   $r_{\mathsf{g}}$,
the TC is insensitive to the spreading factor and exclusion-zone radius.

\begin{figure}[t]
\centering
\vspace{-0.1cm} \includegraphics[width=8.75cm]{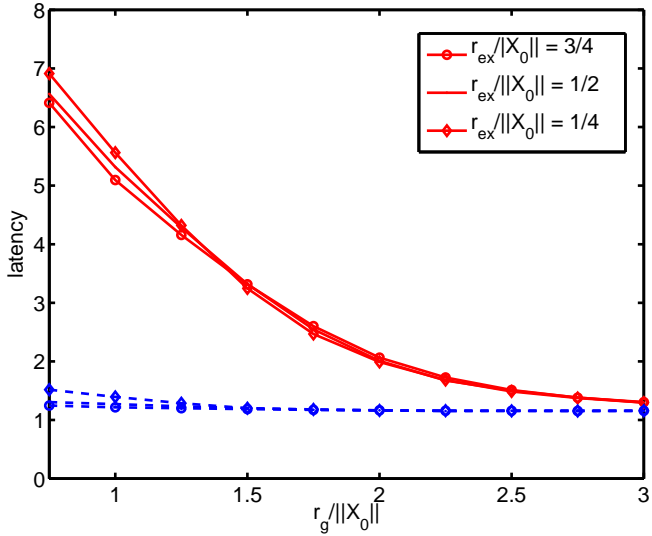} \vspace{-0.65cm}%
\caption{Average latency (in units of time slots) as a function of $r_{\mathsf{g}}/||X_0||$ for several
values of $r_{\mathsf{ex}}/||X_0||$ with $M=30$, $r_\mathsf{net}/||X_0|| = 6$, and $N_{ARQ} = 6$.
. Dashed lines are for spreading ($G_{\mathsf{e}%
}=48$), and solid lines are for no spreading ($G_{\mathsf{e}}=1$). }
\vspace{-0.6cm}
\label{Figure:Latency}
\end{figure}

The average latency is shown in Fig. \ref{Figure:Latency} for $N_\mathsf{ARQ} = 6$.  As the latency is a function of the outage probability, the general trends are similar to those already observed in Fig. \ref{Figure:OutageRg}.  In particular, the unspread network is more sensitive than the spread network to the choice of exclusion-zone radius, and the latency performance of the spread network is better than that of the unspread network.  Except for
small values of guard-zone radius, the latency is relatively insensitive to the radius
of the exclusion zone.  As the exclusion zone radius is limited in practice by the mobility 
constraints of the mobiles, it follows that mobility does not have a significant effect
on latency.

\subsection{Effect of the Reference-Receiver Location}

In a finite network, the average outage probability depends on the location
of the reference receiver. Rather than leaving the reference receiver at the
origin, Table \ref{Table:Table} explores the change in outage probability when the reference
receiver moves from the center of the 
network to its perimeter. The guard and exclusion zones are again normalized 
with respect to $||X_0||$.   Two representative values for each of the
normalized exclusion-zone and guard-zone radii are considered:
$r_{\mathsf{ex}}/||X_0||=\{0,1/2\}$ and $r_{\mathsf{g}}/||X_0||=\{1/2,3/2\}$.
Two different values of path-loss exponent are considered: $\alpha = 3$ and 
$\alpha = 4$.  Both unspread and spread ($G_e = 48$)
networks are considered.  The number of potentially interfering mobiles
are set to $M=30$, and $r_\mathsf{net}/||X_0|| = 6$.  When the
reference receiver is located on the network perimeter, the reference transmitter 
is placed in the direction toward the network center. 

Table \ref{Table:Table} indicates
that the average outage probability on the finite network's perimeter $\bar{\epsilon}_{p}$ 
is considerably less than the average outage probability at the network's center 
$\bar{\epsilon}_{c}$.  This result cannot be predicted by the traditional
infinite-network model, which cannot differentiate between $\bar{\epsilon}%
_{c}$ and $\bar{\epsilon}_{p}$. The reduction in outage probability is more
significant for the unspread network and is less pronounced with increasing $%
G_{\mathsf{e}}$.  Both direct-sequence spreading and CSMA make the average performance less
dependent on the location of the reference receiver inside the network.

\begin{table}[ptb]
\caption{Average outage probability when the reference receiver is at the center ($%
\bar{\protect\epsilon}_{c}$) and on the perimeter ($\bar{\protect\epsilon}%
_{p}$) of the network.\label{Table:Table} }\vspace{-0.3cm} \centering
\par
\begin{tabular}{|c|c|c|c|c|c|}
\hline
$G_\mathsf{e}$ & $\alpha$ & $r_\mathsf{ex}/||X_0||$ & $r_\mathsf{g}/||X_0||$ & $\bar{\epsilon%
}_{c}$ & $\bar{\epsilon}_{p}$ \\ \hline
1  & 3 & 0   & 1/2 & 0.5298 & 0.3056 \\ \cline{4-6}
   &   &     & 3/2 & 0.2324 & 0.1683 \\ \cline{3-6}
   &   & 1/2 & 1/2 & 0.5234 & 0.2592 \\ \cline{4-6}
   &   &     & 3/2 & 0.2256 & 0.1528 \\ \cline{2-6}
   & 4 & 0   & 1/2 & 0.4129 & 0.2388 \\ \cline{4-6}
   &   &     & 3/2 & 0.1453 & 0.1228 \\ \cline{3-6}
   &   & 1/2 & 1/2 & 0.3869 & 0.1774 \\ \cline{4-6}
   &   &     & 3/2 & 0.1313 & 0.1026 \\ \cline{1-6}
48 & 3 & 0   & 1/2 & 0.0644 & 0.0391 \\ \cline{4-6}
   &   &     & 3/2 & 0.0181 & 0.0172 \\ \cline{3-6}
   &   & 1/2 & 1/2 & 0.0308 & 0.0199 \\ \cline{4-6}
   &   &     & 3/2 & 0.0173 & 0.0165 \\ \cline{2-6}
   & 4 & 0   & 1/2 & 0.0842 & 0.0494 \\ \cline{4-6}
   &   &     & 3/2 & 0.0177 & 0.0174 \\ \cline{3-6}
   &   & 1/2 & 1/2 & 0.0335 & 0.0209 \\ \cline{4-6}
   &   &     & 3/2 & 0.0165 & 0.0163 \\ \cline{1-6}
\end{tabular}
\vspace{-0.2cm}
\end{table}

\subsection{Effect of the Number of Potentially Interfering Mobiles}

\begin{figure}[t]
\centering
\vspace{-0.1cm} \includegraphics[width=8.75cm]{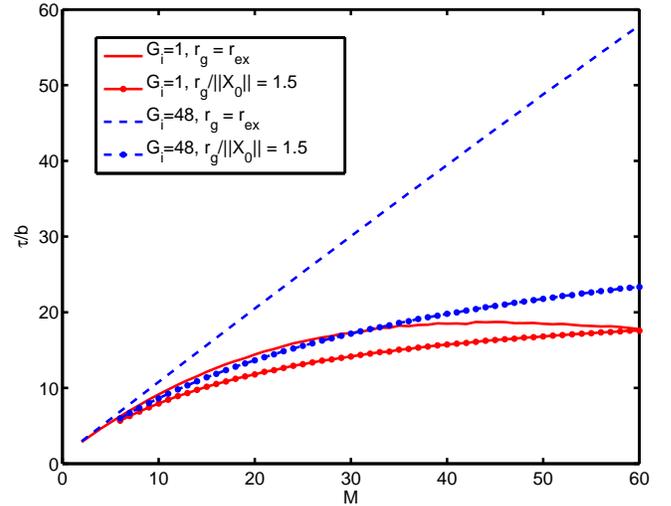} 
\vspace{-0.5cm} 
\caption{Transmission capacity as a function of the number of placed mobiles 
$M$. A normalized exclusion-zone radius of $r_\mathsf{ex}/||X_0|| = 1/2$ is used. 
Performance is
shown both for spreading ($G_\mathsf{e}=48$) and without spreading ($G_%
\mathsf{e}=1$), and both with a CSMA guard zone ($r_\mathsf{g}/||X_0||=3/2$) and
without an additional CSMA guard zone ($r_\mathsf{g}=r_\mathsf{ex}$).  }
\label{Figure:TCvM}
\end{figure}

In the previous figures, a fixed number of potentially interfering mobiles ($%
M=30$) was placed prior to CSMA deactivation. In Fig. \ref{Figure:TCvM}, $M$
is varied from 2 to 60, and the transmission capacity is shown for each
value of $M$ for a normalized exclusion-zone radius $r_\mathsf{ex}/||X_0|| = 1/2$,
$r_\mathsf{net}/||X_0|| = 6$, and both unspread and spread ($G_e = 48$)
networks.  Results are shown for a CSMA guard zone with normalized radius
$r_\mathsf{g}/||X_0||=3/2$ and for a system that uses no additional CSMA
guard zone; i.e., $r_\mathsf{g}/||X_0|| = r_\mathsf{ex}/||X_0|| = 1/2$.

As observed previously for just $M=30$, the transmission capacity of the
DS-CDMA network is higher than that of the unspread network, and using a
CSMA guard zone reduces transmission capacity. The transmission capacity of
the DS-CDMA network increases roughly linearly with $M$ in the absence of a
guard zone, but the increase is sublinear for the unspread network or when
a guard zone is used. At $M=60$, the transmission capacity of the unspread
network is approximately the same with and without a guard zone.

\subsection{Effect of Transmitter Distance}

\begin{figure}[t]
\centering
\includegraphics[width=8.5cm]{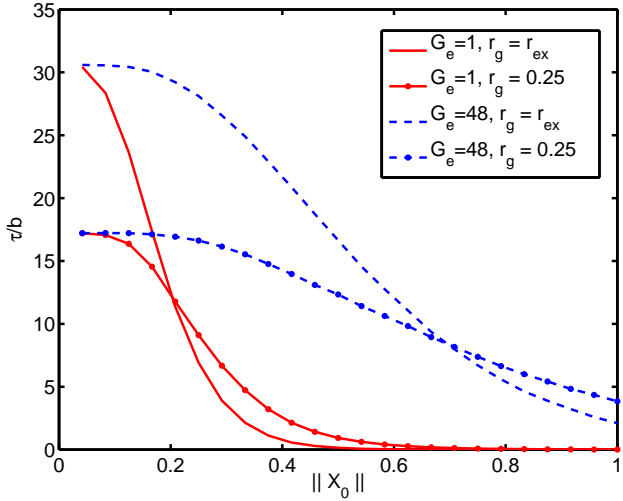} \vspace{-0.25cm} \vspace{-0.1cm}
\caption{Transmission capacity for $r_{\mathsf{ex}}=1/12$ and different
transmit distances $||X_{0}||$. All distances are normalized to the
network radius. }
\label{Figure:MoveTX}
\vspace{-0.4cm}
\end{figure}

The previous examples assume that the distance between the reference
transmitter and receiver is fixed with respect to the network radius.
However, performance will depend on this distance. 
Fig. \ref{Figure:MoveTX} shows $\tau /b,$ the normalized TC as a
function of the distance $||X_{0}||$ between the reference transmitter and
receiver. All distances are normalized to the network radius so that $r_{\mathsf{net}}=1$
The exclusion-zone radius is set to $r_\mathsf{ex} = 1/12$, and both 
unspread and spread ($G_e = 48$) networks are considered. 
Results are shown for a CSMA guard zone with radius
$r_\mathsf{g}=1/4$ and for a system that uses no additional CSMA
guard zone.

It is observed that
increasing the transmission distance reduces the TC due to the increase in
the number of interfering mobiles that are closer to the receiver than the
reference transmitter, but this reduction is more gradual with the spread
system than the unspread one. As 
$||X_0||$
increases, an increased guard zone
alleviates the potential near-far problems. Consequently, the rate of TC
degradation is made more gradual by the use of CSMA, and at sufficiently
large transmitter distances, a system with CSMA outperforms a system without
it.   

The CSMA guard zone decreases transmission capacity at short transmission
distances, but increases it at long distance.  This is because at a long
transmission distance, the received signal power from the far reference
transmitter will be weak, while the received powers from the nearby
interferers will be relatively high.  To overcome this near-far problem, 
nearby interferers need to be deactivated in order for the SINR threshold
to be met.  However, at short distances, the signal power from the nearby
reference transmitter will already be strong enough, 
and deactivating the interferers is unnecessary and harmful to the 
transmission capacity due to a reduction of simultaneous transmissions.

\subsection{Tradeoff Between Spreading Factor and $r_\mathsf{g}$}

The average outage probability depends on both the effective spreading factor 
$G_{\mathsf{e}}$ and guard-zone radius $r_{\mathsf{g}}$. Fig. \ref%
{Figure:OutageConstraint} explores the tradeoff between these two parameters
by plotting the minimum $r_{\mathsf{g}}$ required to achieve an average
outage probability $\bar{\epsilon}\leq 0.1$ for a given value of $G_{\mathsf{%
e}}$. All distances are normalized to the network radius and 
the exclusion-zone radius is set to $r_\mathsf{ex} = 1/12$
The reference transmitter is placed at one of four distances from the reference receiver:
$||X_{0}||\in \{1/2,1/3,1/4,1/6\}$. 
For each $||X_{0}||$, two values of $M$
are considered:, $M=\{30,60\}$. The curves are generated using a Monte Carlo
approach as follows. For each value of $M$, $N=10,000$ networks are realized
by drawing from a uniform clustering process. For each $r_{\mathsf{g}}$,
nodes are deactivated using the Matern thinning policy described in Example
\#2, and new values of $G_{\mathsf{e}}$ are repeatedly picked until one is
found that has a corresponding $\bar{\epsilon}$ within the range $0.0999\leq 
\bar{\epsilon}\leq 0.1010$. It is observed that the required spreading factor
increases as the guard-zone radius decreases until the guard zone coincides
with the exclusion zone.

\begin{figure}[t]
\centering
\vspace{-0.1cm} \includegraphics[width=8.75cm]{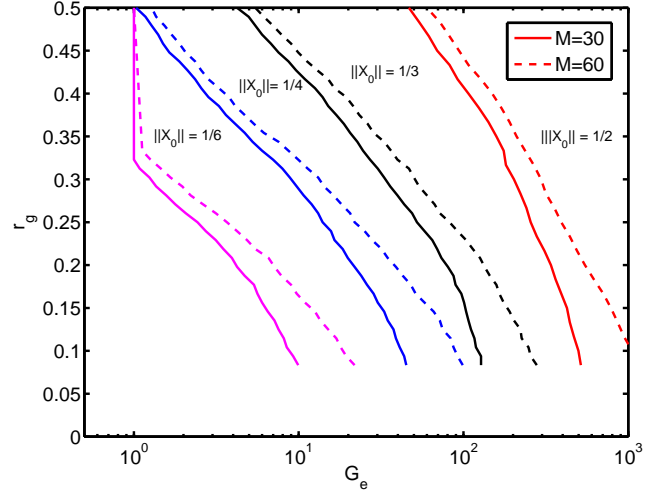} \vspace{%
-0.65cm}\vspace{-0.1cm}
\caption{CSMA guard zone required to achieve an outage probability $\protect%
\bar{\epsilon} \leq 0.1$ when $r_{\mathsf{ex}}=1/12$, the effective processing
gain is $G_{\mathsf{e}},$ and the source transmitter is at four different
distances. All distances are normalized to the network radius.}
\label{Figure:OutageConstraint}
\end{figure}

\begin{figure}[t]
\centering
\vspace{-0.1cm} \includegraphics[width=8.75cm]{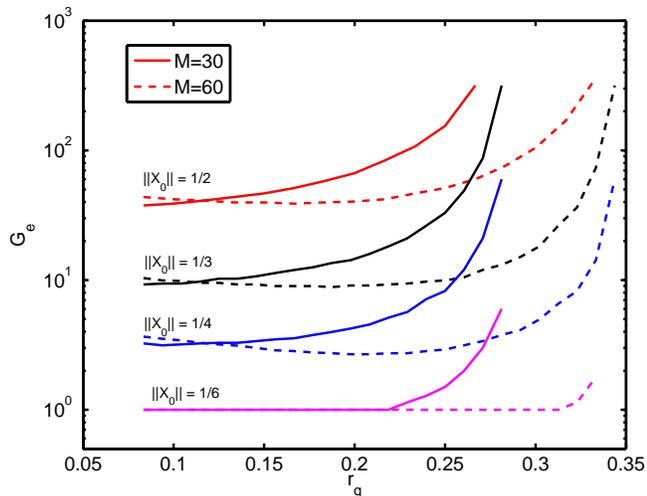} \vspace{-0.65cm}%
\vspace{-0.1cm}
\caption{CSMA guard zone required to achieve a transmission capacity $%
\protect\tau \geq 15b$ when the effective spreading factor is $G_\mathsf{e}$
and the source transmitter is at four different distances.
All distances are normalized to the network radius. }
\label{Figure:TcConstraint}
\end{figure}

As with the outage probability, the transmission capacity depends on both
the effective spreading factor $G_{\mathsf{e}}$ and guard-zone radius $r_{%
\mathsf{g}}$. Fig. \ref{Figure:TcConstraint} explores the tradeoff between
these two parameters by plotting the minimum $G_{\mathsf{e}}$ required to
achieve transmission capacity $\tau \geq 15b$ for a given value of $r_{%
\mathsf{g}}$. As with Fig. \ref{Figure:OutageConstraint}, 
all distances are normalized to the network radius, $r_\mathsf{ex} = 1/12$, the 
reference transmitter is placed at one of four distances, and two values of $M$ are considered. 
The curves are generated using a Monte Carlo approach similar
to the one used to generate Fig. \ref{Figure:OutageConstraint}. The need for
a larger $G_{\mathsf{e}}$ as $r_{\mathsf{g}}$ grows is consistent with the
observation that the transmission capacity generally decreases as $r_{%
\mathsf{g}}$ (see Fig. \ref{Figure:TCvRg}). Although the average outage
probability at the reference receiver decreases with increasing $r_{\mathsf{g}}$%
, the number of interfering mobiles is reduced due to the thinning, and
hence the transmission capacity is reduced.

\section{Conclusions}

\label{Section:Conclusions}

The analysis developed in this paper allows the tradeoffs between exclusion
zones and CSMA guard zones to be explored for DS-CDMA and unspread networks.
The spreading factor and the guard-zone size provide design flexibility in
achieving specified levels of average outage probability and transmission
capacity. The advantage of an exclusion zone over a CSMA guard zone is that
since the network is not thinned, the number of active mobiles remains
constant, and higher transmission capacities can be achieved.  However,
practical constraints of mobility and positioning of mobile terminals can sometimes prevent the use of a significant exclusion zone.  In these cases, a CSMA guard zone might be the only option.  While exclusion zones are advantageous when they can be imposed, when they cannot, a CSMA guard zone still provides performance benefits.

The computational results indicate a number of general tendencies for
networks with exclusion and guard zones. Increases in the size of the
exclusion zone are beneficial, but there are practical limits to this size,
and the transmission capacity associated with larger exclusion zones
diminishes quickly with increasing guard zones. The CSMA guard zone is much
more useful in networks with unspread systems than those with
direct-sequence spreading. Both spreading and a CSMA guard zone make the
outage probability less dependent on the location of the receiver within the
network. The spreading factor required for a specified outage probability
increases with decreases in the guard zone until the guard zone coincides
with the exclusion zone. The rate of transmission-capacity degradation as
the transmitter distance increases is made more gradual by the use of CSMA,
and at sufficiently large transmitter distances, a system with CSMA
outperforms a system without it. Although both the average outage probability
and the latency of a receiver decrease with increasing guard zones, the number of
interfering mobiles is reduced due to the thinning, and hence the
transmission capacity is reduced.


\balance

\bibliographystyle{ieeetr}

\ifpdf
  \begin{IEEEbiography}{Don Torrieri}
\else
  \begin{IEEEbiography}[{\includegraphics[width=1in,height=1.25in,clip,keepaspectratio]{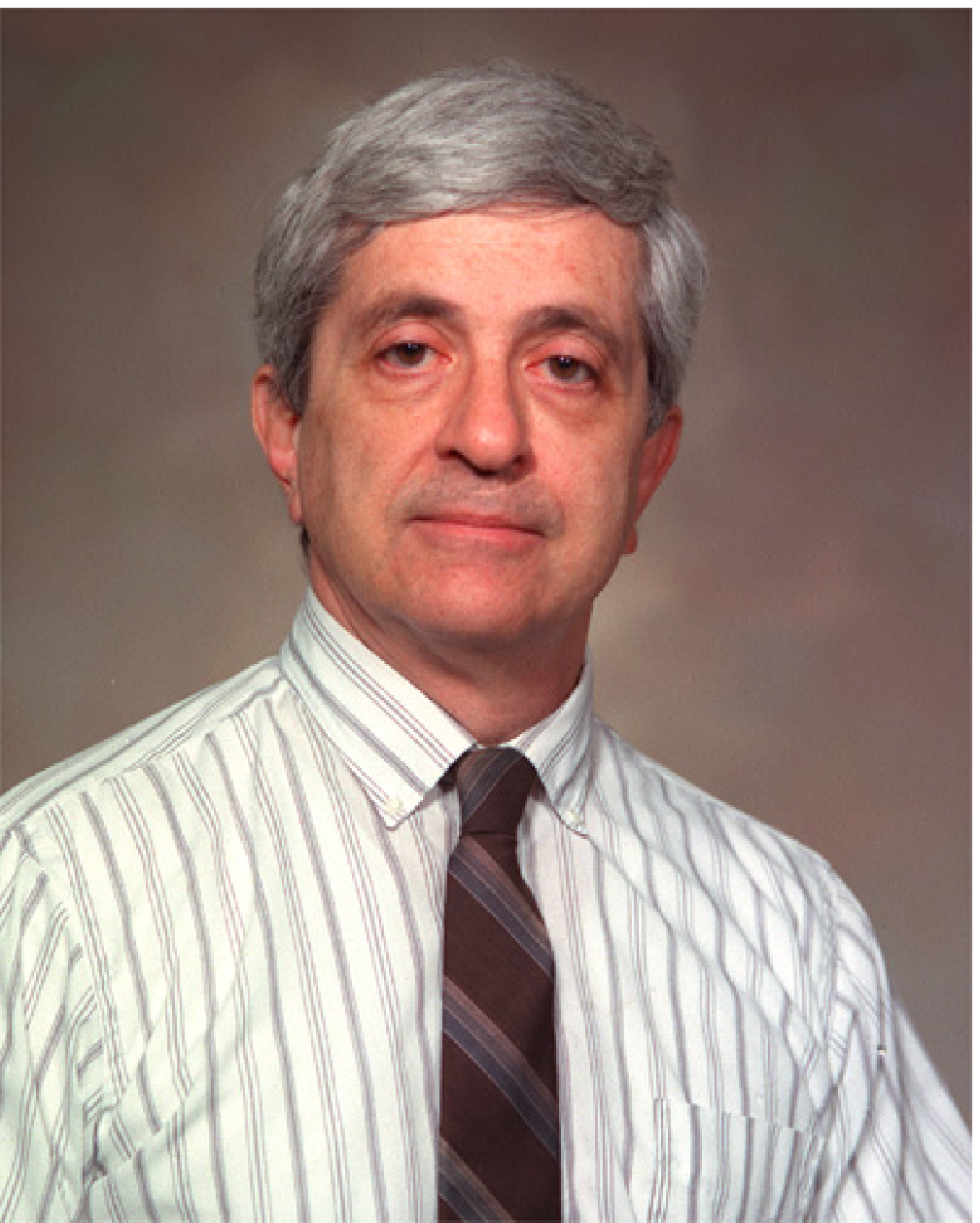}}]{Don Torrieri}
\fi
 is a research engineer and Fellow of the US Army Research Laboratory. His primary research interests are communication systems, adaptive arrays, and signal processing. He received the Ph. D. degree from the University of Maryland. He is the author of many articles and several books including {\em Principles of Spread-Spectrum Communication Systems}, 2nd ed. (Springer, 2011). He teaches graduate courses at Johns Hopkins University and has taught many short courses. In 2004, he received the Military Communications Conference achievement award for sustained contributions to the field.
\end{IEEEbiography}

\ifpdf
  \begin{IEEEbiography}{Matthew C. Valenti}
\else
  \begin{IEEEbiography}[{\includegraphics[width=1in,height=1.25in,clip,keepaspectratio]{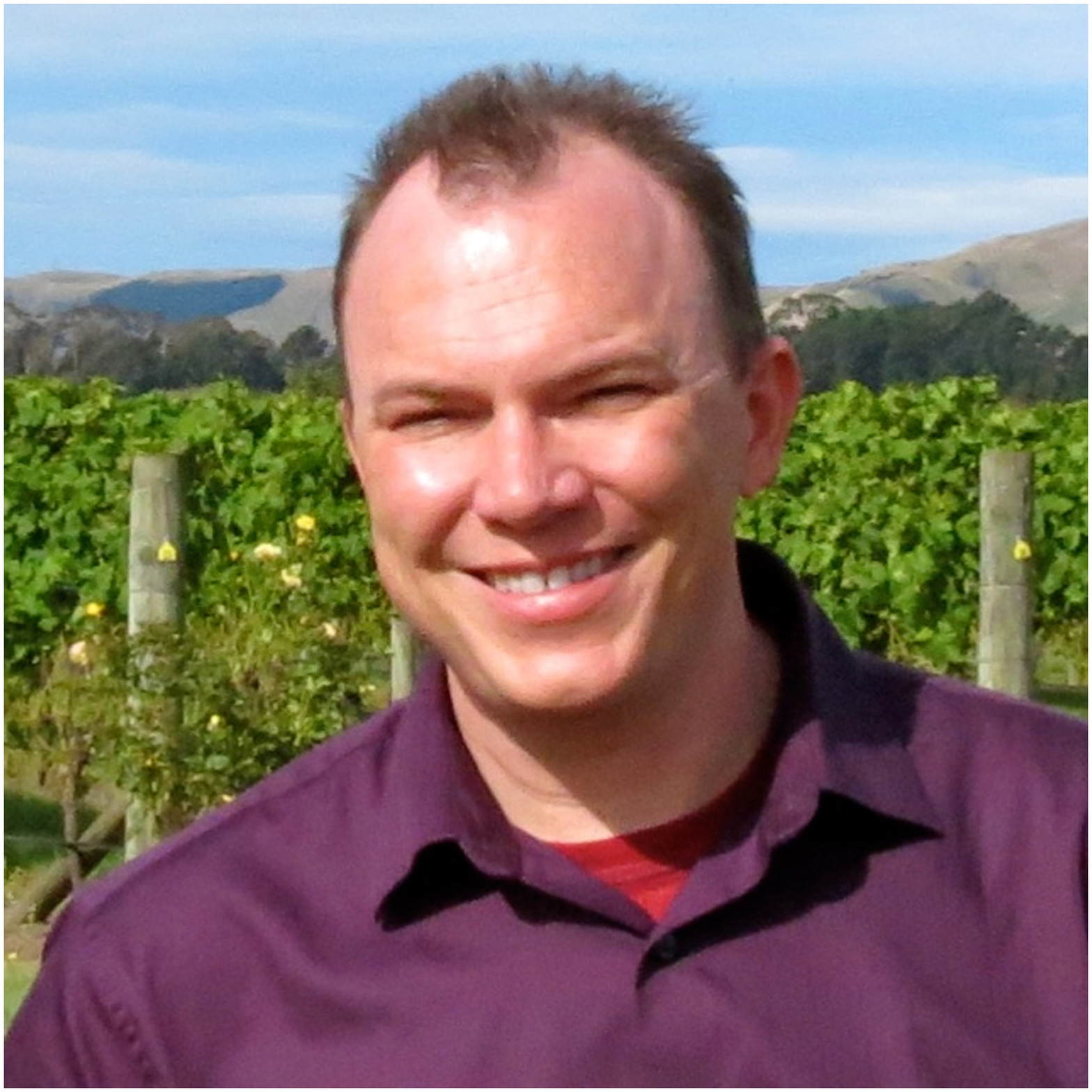}}]{Matthew C. Valenti}
\fi
is a Professor in Lane Department of Computer Science and Electrical Engineering at West Virginia University. He holds BS and Ph.D. degrees in Electrical Engineering from Virginia Tech and a MS in Electrical Engineering from the Johns Hopkins University. From 1992 to 1995 he was an electronics engineer at the US Naval Research Laboratory.  He serves as an associate editor for {\em IEEE Wireless Communications Letters} and as Vice Chair of the Technical Program Committee for Globecom-2013.  Previously, he has served as a track or symposium co-chair for VTC-Fall-2007, ICC-2009, Milcom-2010, ICC-2011, and Milcom-2012, and has served as an editor for {\em IEEE Transactions on Wireless Communications} and {\em IEEE Transactions on Vehicular Technology}. His research interests are in the areas of communication theory, error correction coding, applied information theory, wireless networks, simulation, and secure high-performance computing.  His research is funded by the NSF and DoD.  He is registered as a Professional Engineer in the State of West Virginia.
\end{IEEEbiography}

\end{document}